\documentclass[11pt,letter]{article}
\usepackage{graphicx,amscd,amsmath,amssymb,amsfonts,verbatim}

\usepackage{hyperref}
\hypersetup{colorlinks = true, linkcolor  = blue}

\usepackage[utf8]{inputenc}
\usepackage{graphicx}
\usepackage{amssymb}
\usepackage{amsmath}
\usepackage{caption}
\usepackage{subcaption}

\usepackage{tikz-cd}

\def\keywords#1{{\bf Keywords}\\{#1}}  %{\begin{center}{\bf Keywords}\\{#1}\end{center}}

\usepackage{hyperref}

\usepackage[shortlabels]{enumitem}

\usepackage{mathrsfs}
\usepackage{bbm}
\usepackage[all]{xy}
\usepackage[authoryear]{natbib}
\usepackage{authblk}
\usepackage{fancyhdr}

\usepackage{amsthm}
\numberwithin{equation}{section}

\DeclareMathAlphabet{\mathpzc}{OT1}{pzc}{m}{it}

\date{}

\theoremstyle{definition}

\newtheorem{remark}{Remark}[section]
\newtheorem{theorem}{Theorem}[section]

\begin{document}

{
\title{Estimation for the Cox Model with Biased Sampling Data via Risk Set Sampling}
\author{}  
%\if 1 \blind   
\author{Omidali Aghababaei Jazi\\
Department of Mathematical and Computational Sciences \\
University of Toronto Mississauga.} %\fi
\maketitle
}

\if 1 \blind
{
  \bigskip
  \bigskip

  \bigskip
  \begin{center}
    {\LARGE\bf Title}
\end{center}
  \medskip
} \fi

%\affil{Department of Mathematics and Statistics, \\ McGill University}

%\maketitle

\pagestyle{plain}

%\bigskip
\section*{Abstract}
Prevalent cohort sampling is commonly used to   study the natural history of a  disease  when the disease is rare or
it usually takes a long time to observe the failure event.  It is known, however, that the collected sample in this situation is  
not representative of the target population  
which in turn  leads to biased sample risk sets.
In addition, when survival times are subject to censoring,  the censoring mechanism is informative.  In this paper, 
I propose a pseudo-partial likelihood estimation method   for  estimating parameters in 
the Cox proportional  hazards model with right-censored and biased sampling data by adjusting sample risk sets. I study the asymptotic properties of 
 the resulting  estimator  and conduct a simulation study   to   illustrate  its finite sample  performance of the proposed method.  
 I also use the proposed method to analyze a set of   HIV/AIDS data. 

\noindent \keywords{Biased Sampling;  Cox Regression Model,  Right-censoring;   Risk Set Sampling. }

%%%%%%%%%%%%%%%%%%%%%%%%%%%%%%%%%%%%%%%%%%%%%%%%%%%%%%%%%%%%%%%%%%%%%%%%
%%%%%%%%%%%%%%%%%%%%%%%%%%%%%%%%%%%%%%%%%%%%%%%%%%%%%%%%%%%%%%%%%%%%%%%%

\section{Introduction}
\label{intro}
Incident and prevalent cohort sampling designs are commonly used to study   the natural history of diseases.  An incident cohort sampling  design 
is the gold standard in survival analysis. But,  it can be infeasible in practice when  the disease under study  is rare or it  usually takes a long time
 to observe the failure event.  \\
A prevalent cohort sampling design  is a viable alternative that follows forward  subjects  who have experienced an initiating event before the recruitment. 
The design, however, poses two challenges.   Firstly,  when the collected sample are subject to censoring, the censoring mechanism 
is informative \citep{asgharian2005}. 
Secondly, the collected sample are not a representative of the target population since a patient has to survive until the 
recruitment time  to have a chance to be included  in the study. Therefore,  the collected sample is biased toward long survival times.
and as such sampling is biased. Under this situation, sample risk sets  do not form a random sample from the risk population 
and the standard methods for  estimating parameters become inappropriate.  \\
Under biased sampling design, the observed survival times are said to be left-truncated.   
When the incidence rate of the onset of the disease  remains almost constant 
over time (the stationarity assumption),  the type of bias  is known as length-bias. 
The stationarity assumption is closely tied to uniform truncation  distribution \citep{asgharian2003}.   
In fact, under some mild conditions, the stationarity assumption holds if and only if the truncation time is uniformly distributed  \citep{masoud2006}. \\
There has been a rising interest in nonparametric and semiparametric estimation  for prevalent cohort data  over the last two decades.
\citet{jacobo2020}   summarized  nonparametric estimation methods of the survival function 
when the distribution of  truncation times  is either partially or completely known. 
They developed  two methods of estimating for both the truncation and the survival  distributions under a semiparametric truncation model 
in which the truncation variable is assumed to have  a certain parametric distribution.  
\citet{shenetal2017} also provided a  thorough review of   the nonparametric and  semiparametric  estimation methods  for  right-censored 
length-biased data. In particular, the estimation methods  for the Cox proportional hazards (PH) model with  right-censored length-biased data  
can be  classified into  the weighted estimating equation   and  likelihood-based approaches.  The former uses some weight functions 
to adjust length-biased data \citep{qinshen2010}. The latter, on the other hand, uses the conditional likelihood of observed 
survival data given the truncation times \citep{huangqin2012}.
With right-censored and left-truncated data,  \citet{wangetal1993} proposed the partial likelihood method for estimation under the Cox PH model. 
The partial likelihood function  is similar to the that of classical survival data, except for the structure of the risk set. 
The authors showed that the maximum  partial likelihood estimator is  asymptotically as efficient as an estimator obtained from the truncation 
likelihood function  when the  truncation distribution is unknown. However, when the truncation  distribution 
is known, the  partial likelihood may lead  to a loss of efficiency.  
\citet{wang1996} proposed a pseudo-partial likelihood method for estimating parameters in the Cox PH model for
length-biased data in the absence of right-censoring  by correcting the bias of  sample risk sets. \citet{tsai2009}  developed  a pseudo-partial likelihood 
method for   the Cox PH model  with biased sampling data by embedding the  data into left-truncated data which in turn yields a more  efficient estimator.  \\  
In this article, following  the idea of risk set sampling \citep{wang1996}, I propose a pseudo-partial likelihood method  under the Cox PH model 
with  right-censored and  biased sampling data   by correcting the bias of  sample risk sets.  
%The extension has two folds. Firstly, I consider general biased sampling data and, secondly, in the presence of right-censoring. 
I study the large sample properties of the resulting estimator, conduct a simulation study to confirm its finite  sample performance, and apply the method 
to analyze a  dataset from the  HIV-infection and AIDS. \\
The rest of this article is organized as follows. In Section 2,  some notation, the model,  and the likelihood function of the observed data are introduced.  
In Section 3, I first introduce the pseudo-partial likelihood method   and then discuss  the large sample properties of the resulting estimator. In Section 4, I conduct
a simulation study  to confirm  the finite sample performance of the proposed method. 
I also apply the proposed method to analyze a  a set of   HIV/AIDS data.
Section 5 provides some discussion and   closing remarks. The Appendix includes proofs and  other   details.

\section{Notation and Preliminaries}
\label{sec:examples}
\begin{figure}
\centering
\includegraphics[scale=0.25]{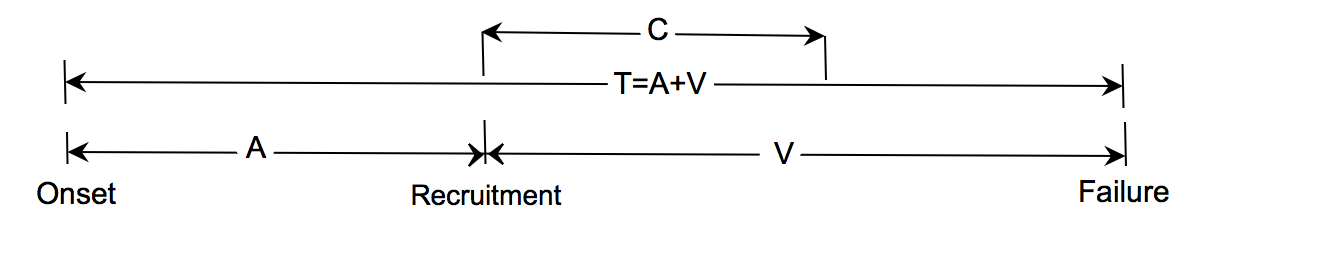}
\caption{Positions of variables in a prevalent cohort study.}
\label{fig:1}
\end{figure}

Let $\tilde T$  and $\tilde A$ be the survival time and     the truncation time, respectively 
and $\tilde {\textbf Z}$ be the corresponding  $p \times 1$ vector of  covariates  in a  target population. 
Let  $f_U(t|\textbf z)$ and  $S_U(t|\textbf z)$ be the  conditional probability density function and   the survival function  
of  $\tilde T$ given $\tilde {\textbf Z}=\textbf z$,  respectively. Let  further   $g_{\theta}(a)$ and $G_{\theta} (a)$ be    
the probability density   function and the distribution function  of $\tilde A$ with a known parameter $\theta$, respectively. \\
Under prevalent cohort sampling, the collected sample are not representative of the target population. I therefore use different variables
to distinguish them  from their counterparts in  the target population. 
Let  $T$   represent the survival time, $A$ be the observed   truncation time which is also called the backward recurrence time,  and $\textbf Z$
be  the covariate  vector  in a prevalent cohort   sample.   Let  $V$  be the time from the recruitment until failure, 
also known as the forward recurrence time; i.e. $V=T-A$  (see Figure \ref{fig:1}).  \\
Survival times  in a prevalent cohort may  be subject to right censoring.
Let $C$ represent the time from  the recruitment to censoring, called the residual censoring time. The total censoring time  is therefore $\tilde C= A+C$.
Let further $Y = \min (T, \tilde C)$ and $\Delta  = I( {T \le {\tilde C}})$  denote  the observed
survival time and the censoring indicator.  The  observed  data   $(a_{i}, y_{i},  \delta_{i}, \textbf{z}_{i})$ for $i=1,2, \cdots, n$. 
are then  assumed  to be independent and identically distributed realizations  of $(A,Y, \Delta, \textbf Z)$.
\\
Suppose that the hazard of failure times in the target population  follows  the Cox PH model.
\begin{align}
\label{eq: 2.1}
h( {t |  \tilde{\textbf Z} ={\textbf z}}) = {h}(t)\exp ({\pmb\beta'}{\textbf z} ),
\end{align}
where $h(t)$ is an unspecified baseline hazard function and $\pmb \beta$ is 
a $p \times 1$ vector of parameters. \\
Under  prevalent cohort design,  the joint distribution of  $(A,T,\textbf Z)$ is  the same as that  of   $(\tilde A,\tilde T,  \tilde {\textbf Z})$ 
given $\tilde T \ge \tilde A$. Then it is not hard  to show  that     the   probability  density function of $(A,T)$ given  ${\textbf Z}={\textbf z}$  is given by
\begin{align}
\label{eq: 2.2}
{f_{(A,T)}}\left( {a,t\left| {\textbf z} \right.} \right) = \begin{cases}
\frac{g_{\theta}(a){f_U(t\left| {\textbf z} \right.)}}{{\mu_G ({\textbf z})}}  &\mbox{ $t >a$}\\         
0 &\mbox{Otherwise,}
\end{cases}
\end{align}
where   $g_{\theta}(t)$ is the  probability density function  of $\tilde A$ with some known  parameter $\theta$ and
$$\mu_G ({\textbf z}) = E [ {\tilde T| {\tilde {\textbf Z} = {\textbf z}}}]=\int_0^{\infty} S_U (t | \textbf z)g_{\theta}(t)dt.$$
It follows from  the joint probability   density function  \eqref{eq: 2.2}  and  the Cox PH model \eqref{eq: 2.1} that 
the full likelihood function of the collected  sample is proportional to
\begin{align}
\label{eq: 2.4}
L_F   &\propto {\prod\limits_{i = 1}^n {\left\{ {\frac{{{f_U}({y_i}|{{\textbf z}_i})g_{\theta}({a_i})}}{{{\mu_G}({{\textbf z}_i})}}} 
\right\}}^{{\delta _i}}}{\left\{  {\frac{{{S_U}({y_i}|{{\textbf z}_i})g_{\theta}({a_i})}}{{{\mu _G}({{\textbf z}_i})}}} \right\}^{1 - {\delta _i}}}  \nonumber \\
&=\prod\limits_{i = 1}^n {{{\left\{ {{h}({y_i}){\exp ({{{\pmb \beta'}}{{\textbf z}_i}})}} \right\}}^{{\delta _i}}}\frac{{\exp \left\{ 
{ - {H({y_i}){\exp ({{{\pmb\beta'}}{{\textbf z}_i}})}} } \right\}}}{{\mu_G ({{\textbf z}_i})}}}g_{\theta}({a_i}),
\end{align}
where $H(.)$ is the cumulative baseline  hazard function.  \\
Although direct  maximization of    equation \eqref{eq: 2.4}  with respect to $(\pmb\beta, H(.))$ yiels most efficient estimators of the parameters,  
it  is numerically intractable especially  for large sample sizes  since it involves  the integral of the  nonparametric $H (.)$ 
in a complicated way. I therefore propose a pseudo-partial likelihood method by correcting the bias of sample risk sets.

\section{Estimation and Asymptotic}
\label{est}
Under classical   survival data,   the inference based on the partial likelihood   for  the Cox PH model
is driven by risk sets just prior vto the failure times.   For prevalent cohort data, by contrast,   a subject in the target population may experience 
the failure event before the start of the study and hence sample  risk sets  do not form a random sample from the population risk sets 
and they require some adjustment.

\subsection{Pseudo-partial Likelihood Method}
Let $m$ be the number of subjects that are observed to fail  at  
times ${t_{(1)}} < ... < {t_{(m)}}$  ($m \le n$) 
and  ${\mathcal R}_i = \{ j:{y_j} \ge {t_{(i)}},{\delta _j} = 1\}$ for $i=1,..., m$, be  the sample risk  set of uncensored subjects.
For  subject $j$  in $\mathcal R_i$, define  the indicator variable  ${\xi_j}(u)$  that  assumes   1 with
probability ${\Omega _C}({u})/{\Omega _C}({y_j})$  and 0 with probability $1-{\Omega _C}({u})/{\Omega _C}({y_j})$ for  $0 < u \le y_j$
where ${\Omega_C}(y)=\int_0^y {{S_C}(t)}g_{\theta}(t)dt$. The bias-adjusted risk set is then defined   by
\begin{align}
\label{eq: 2.3.3}
\tilde {\mathcal R}_i = \{ j:{y_j} \ge {t_{(i)}} , {\delta_j} = 1 , {\xi_j}({t_{(i)}}) = 1\}.
\end{align}
It can be shown that  the subjects in $\tilde {\mathcal R}_i$ have the population risk structure. To this end,  first note that
given that $C$ is independent of $(A,T)$ and the distribution of $C$ does not depend on   covariates,  by the joint probability density function \eqref{eq: 2.2}, 
the conditional  probability of observing uncensored data given $\textbf Z= \textbf z$  is 
\begin{align}
\label{eq: 2.3.2}
P(Y = y,\delta  = 1| {\textbf z}) = \int_0^y {P(T = y, A = a | {\textbf z})P(C \ge y - a)da}
= \frac{{{f_U}(y|\textbf z)}}{\mu_G(\textbf z) }{\Omega_C}(y).
\end{align}
It follows from  the indicator  function and equation \eqref{eq: 2.3.2} that for subjects in $\tilde {\mathcal R}_i$, 
sampling $\tilde {\mathcal R}_i$  is equivalent to the sampling conditional probability  
of $\tilde T$   for $\tilde T \ge t_{(i)}$ 
because 
\begin{align}
&pr({y_j}\left| {{\xi_j}({t_{(i)}}) = 1,{Y_j}} \right. \ge {t_{(i)}},{\delta _j} = 1,{\textbf z_j})  \nonumber \\
&= \frac{{{pr}( {\xi_j}({t_{(i)}}) = 1\left| {{y_j},{Y_j}} \right. \ge {t_{(i)}},{\delta _j} = 1,{\textbf z_j})
pr({Y_j=y_j},{\delta _j} = 1\left| {{Y_j} \ge {t_{(i)}},{\textbf z_j}} \right.) }}{{\int_0^{\infty} {{pr}( {\xi _j}({t_{(i)}})
= 1\left| {y,{Y_j}} \right. \ge {t_{(i)}},{\delta_j} = 1,{\textbf z_j}) pr(Y_j=y,{\delta_j} = 1\left| {Y_j \ge {t_{(i)}},{\textbf z_j}} \right.) dy} }}\nonumber \\
&= \frac{{{\{ {\Omega _C}({t_{(i)}})/{\Omega _C}({y_j})\} }{\Omega _C}({y_j}){f_U}({y_j}\left| {{\textbf z_j}}
\right.)I({y_j} \ge {t_{(i)}})/\mu({\textbf z}_j) }}{{\int_0^{\infty} {{\{ {\Omega _C}({t_{(i)}})/{\omega _C}(y)\} }{\Omega _C}(y){f_U}(y\left|
{{\textbf z_j}} \right.)I(y \ge {t_{(i)}})/\mu({\textbf z}_j)dy} }}\nonumber \\
 &= \frac{{{f_U}({y_j}\left| {{\textbf z_j}} \right.)I({y_j} \ge {t_{(i)}})}}{{\int_0^{\infty} {{f_U}(y\left| {{\textbf z_j}} \right.)I(y \ge {t_{(i)}})dy} }}
= pr({y_j}\left| {\tilde T \ge {t_{(i)}}}, \right.{\textbf z_j}).   
\end{align}
This implies that 
\begin{align}
\label{eq: 2.5.1}
&pr({\textrm{subject  (j) fails at}} \ t_{(i)} |  {\tilde{\mathcal R}_i} )
\approx \frac{{{h_{(j)}}(t_{(i)}| {\textbf z}_{(j)})}}{{\sum\limits_{j \in {\tilde{\mathcal R}_i}} {{h_j}(t_{(i)} | {\textbf z}_{j})} }}   
= \frac{{\exp ({{\pmb \beta'}}{{\textbf z}_{(j)}})}}{{\sum\limits_{j \in {{\tilde {\mathcal R}}_i}} {\exp ({{\pmb \beta'}}{{\textbf z}_j})} }}
\end{align}
Therefore  a  pseudo-partial likelihood function is given by
\begin{align}
\label{eq: 2.3.4}
L(\pmb{\beta}) = \prod\limits_{i = 1}^m {\frac{{\exp ({\pmb \beta' }{\textbf z_{(i)}}) }}{{\sum\limits_{j \in {\tilde {\mathcal R}_i}}
{\exp ({\pmb \beta'}{\textbf z_j}) } }}}.
\end{align}
The maximum pseudo-partial likelihood  estimator  is obtained by maximizing   the likelihood function
after inserting $\hat {\Omega}_C (t)$  from  the Kaplan-Meier estimate of $S_C(t)$. 
Note that   the  (normalized) pseudo-partial score function becomes 
\begin{align}
\label{eq: 2.3.6}
U_n(\pmb{\beta})= \frac{1}{n} \sum\limits_{i = 1}^n {U_i}  = \frac{1}{n} \sum\limits_{i = 1}^n {\delta_i}  
\left[ {{\textbf  z_i} - \frac{{\sum\limits_{j \in {\tilde {\mathcal R}_i}} {{\textbf  z_j}
\exp ({\pmb \beta'}{\textbf  z_j})} }}{{\sum\limits_{j \in {\tilde {\mathcal R}_i}} {\exp ({\pmb \beta'}{\textbf  z_j})} }}} \right].
\end{align}
\begin{remark}
\label{rem: 1}
It is not hard to show that $V_n (\pmb \beta)$ below is an extension of the weighted estimating function  \citep{qinshen2010} for  
length-biased sampling data to biased sampling data. 
\begin{align}
\label{eq: 11}
V_n (\pmb \beta) = \frac{1}{n} \sum\limits_{i = 1}^n V_i=  \sum\limits_{i = 1}^n {  {\delta _i}
\left[ {{\textbf  z_i} - \frac{{\sum\limits_{j = 1}^n {I({y_j} \ge {y_i}){\delta _j}{\textbf z_j}\exp ({\pmb \beta' }{\textbf  z_j}){{\{ {\Omega _C}({y_j})\} }^{ - 1}}} 
}}{{\sum\limits_{j = 1}^n {I({y_j} \ge {y_i}){\delta _j}\exp ({\pmb \beta' }{\textbf  z_j}){{\{ {\Omega _C}({y_j})\} }^{ - 1}}} }}} \right]}
\end{align}
Because 
$E\left[{B_{li}}(y)|y_i \right]= \left[\Omega_C(y)/\Omega_C(y_i)\right]{
\zeta_i}(y)$, 
the difference between $V_i$ and  $U_i$ is the replacement of $\xi_j (Y_i)$ in $U_i$ by the expected value of $\xi_j (Y_i)$ for 
given $(Y_i, Y_j)$.
I show  in the Appendix that   $n^{1/2}  U_n (\pmb \beta)$ and $n^{1/2}  V_n (\pmb \beta)$  are asymptotically equivalent. 
\end{remark}
\begin{remark}
The pseudo-partial likelihood
function \eqref{eq: 2.3.4}  is based on the unbiased risk sets  $\tilde {\mathcal R}_i$  
which, conditional on ${\mathcal R}_i$, is  a random subset of ${\mathcal R}_i$ and ranges  
from  1 to $|{\mathcal R}_i|$. When $|\tilde {\mathcal R}_i|=1$, i.e., 
$\tilde {\mathcal R}_i = \{ i \}$,    
the   pseudo-partial score 
function \eqref{eq: 2.3.6} makes no contribution to the estimation of $\pmb \beta$.  
As suggested by   \citet{wang1996},    the statistical variation   from the risk set  sampling  can be reduced by 
replicating  the   method and 
estimating $\pmb \beta$  by the average of the resulting estimators.  
As the author pointed out, while  the replication procedure does not increase the asymptotic efficiency for estimating $\pmb \beta$,   
it can improve estimation when sample size is small or moderate. \\
Let $l$ be the repetition number and $\xi_{lj} (y_i)$ be the indicators from the $l$th repetition
for $l=1,...,L$. Let $\hat {\pmb \beta}_l$ denote the $p \times 1$ estimator obtained  in the $l$th repetition, 
and $\bar \beta_L=L^{-1}\sum\nolimits_{l=1}^L \hat{\pmb \beta}_l$. While the replication procedure does not increase the asymptotic efficiency 
for estimating $\pmb \beta$,   it is expected to improve estimation when the sample size $n$ is small or moderate. 
\end{remark}
\begin{remark}
The Breslow's estimator \citep[Discussion]{cox1972} for estimating cumulative baseline hazard function with
right-censored survival data   can be adapted to right-censored general biased sampling data. \\
Let  $y_{(i)}$ be  the $i$th distinct  uncensored  data and  $d_i$ be its distance  from  the  nearest $y_{(k)}$ on the left side.  
We assume that  the baseline hazard function  $h_0(y)$ is piecewise constant in $y_{(k)} < y < y_{(i)}$   and let $h_{0i}=h_0(y_{(i)})d_i$.   
By   heuristic argument and  the relationship 
$$1 = {\sum\limits_{j:{y_j} \ge {y_{(i)}}} {{{\left[ {{h_{0i}} {\left\{ {{\Omega _C}({y_{(i)}})/{\Omega _C}({y_j})} \right\}}
\exp ({{\pmb \beta'}}{\textbf z_j})} \right]}}}} 
$$
Substituting  $\hat{\pmb \beta}$ into the above equation,   an estimate of  $h_{0i}$ is %  the MAPLE,  
$${\hat h}_{0i} =  \frac{1}{{\sum\limits_{j:{y_j} \ge {y_{(i)}}} {{{\left[ {\{ {\Omega _C}({y_{(i)}})/{\Omega _C}({y_j})\} 
\exp ({\hat{\pmb \beta'}}{\textbf z_j})} \right]}}} }}$$  % ^{{\delta _i}}  {{\left[ {\exp ({\hat{\pmb \beta}_1^T}{\textbf Z_j})} \right]}^{1 - {\delta _i}}}
Then an estimator for  the cumulative baseline hazard function becomes
$$\widehat{H_0}(t) = \sum\limits_{i: y_{(i)} < t} { {\hat h_{0i}} }$$
\end{remark}

\subsection{Asymptotic  Properties}
Let $N_i(t)=I(y_i \le t, \delta_i=1)$ 
represent  the counting process of  observed failure times. For  each $(l,y)$ define
\begin{align}
\label{eq: 3.19}
{B_{lj}}(y)=
\begin{cases}
{\xi_{lj}(y)} &\mbox{if} \ {y_j} \ge y \  \mbox{and} \ \delta_j=1, \\
0  & \mbox{Otherwise}
\end{cases}
\end{align}
Note that $E\left[{B_{li}}(y)|Y_i \right]= \left[\Omega_C(y)/\Omega_C(Y_i)\right]{\zeta_i}(y)$ 
where ${\zeta_i}(y)=I({y_i} \ge y,{\delta_i} = 1)$ is  the at-risk function for  uncensored observations. 
Define  ${S_{lB}^{(k)}}({\pmb \beta},y) = n^{-1}\sum\nolimits_{j = 1}^n {B_{lj}(y){\textbf z}_j^{ \otimes k}\exp({\pmb \beta'}}{\textbf z_j})$  
for $k=0,1,2$,   with   expectations  ${s^{(k)}}({\pmb \beta},y)$   where for a column vector   $\textbf a$,  ${\textbf a}^ {\otimes 2} = {\textbf a}{\textbf  a'}$, 
${\textbf a}^{\otimes 1}= {\textbf a}$, and ${\textbf  a}^{\otimes 0}=1$.    Then the  pseudo-partial score function \eqref{eq: 2.3.6} can be  expressed as
\begin{align}
\label{eq: 3.22}
U_n(\pmb \beta)= \frac{1}{n}\sum\limits_{i = 1}^n U_{li}  = \frac{1}{n}\sum\limits_{i = 1}^n  {\delta _i}{\left[ {{{\textbf z}_i} 
- \frac{{S_{lB}^{(1)}(\pmb \beta ,{Y_i})}}{{S_{lB}^{(0)}(\pmb \beta ,{y_i})}}} \right]}
\end{align}
For $i=1,...,n$,  define    stochastic process   
\begin{align}
\label{eq: A3.1}
{M_{i}}(t) = {N_i}(t) - \int_0^t \{ {\Omega _C}(y)/{\Omega _C}({y_i})\} {\zeta_i}(y) h(y| \textbf z_i)dy.   
\end{align}
The stochastic process can be interpreted as the difference between the observed number of events and the expected 
number of events under the assumed model until time $t$. When model \eqref{eq: 2.1} is correctly specified, 
this becomes a mean zero stochastic process because
\begin{align}
E\left[ {{M_{i}}(t)} \right] &= E\left[ E[{{N_i}(t)}|{\textbf Z}_i]  \right] 
- E\left[ {\int_0^t {h(y\left| {{{\textbf Z}_i}} \right.) {\Omega _C}(y) E \left[ {\zeta_i}(y)  /{\Omega _C}({Y_i}) |  {\textbf Z}_i \right]dy} } \right]  \nonumber \\
&= E\left[ {{\int_0^t {\frac{{{f_U}(y\left| {{{\textbf Z}_i}} \right.)}}{{\mu_G ({{\textbf Z}_i})}}{\Omega_C}(y)dy} } } \right]    
- E\left[ {\int_0^t {h(y\left| {{{\textbf Z}_i}} \right.){\Omega _C}(y)\frac{{{S_U}(y\left| {{{\textbf Z}_i}} \right.)}}{{\mu_G ({{\textbf Z}_i})}}dy} } \right] \nonumber \\
&= 0 \nonumber 
\end{align}
Therefore when $S_C$ is known,  $U_n (\pmb \beta)$ can be asymptotically represented by the following  independent and identically 
distributed summation of the mean zero stochastic processes.  
$$U_n(\pmb \beta)= \frac{1}{n} \sum\limits_{i=1}^n \int_0^{\tau} \left \{ \textbf z_i 
-  \frac{s^{(1)}({\pmb \beta},v)}{s^{(0)}({\pmb \beta},v)} \right \}dM_i (v)+ o_P(n^{-1/2})
$$
When $S_C$ is unknown, it can be replaced  by its consistent Kaplan-Meier estimator $\hat S_C$.  
Let $\tilde S_{lB}^{(k)}({\pmb \beta},v)$ be an estimate of $S_{lB}^{(k)}({\pmb \beta},v)$ after inserting $\hat S_C$. Then  
\begin{align}
\label{eq: 15}
\tilde U_n (\pmb \beta) = \frac{1}{n} \sum\limits_{i=1}^n \int_{0}^{\tau} \left \{ \textbf z_i 
-  \frac{\tilde S_{lB}^{(1)}({\pmb \beta},v)}{\tilde S_{lB}^{(0)}({\pmb \beta},v)} \right \}dN_i (v)
\end{align}
I  show in the Appendix  that under the regularity conditions, $n^{1/2} \tilde U_n (\pmb \beta)$ converges weakly to a  mean zero  Gaussian process.  \\
Let $\Gamma_n (\pmb \beta)= \partial {\tilde U}_n (\pmb \beta)/\partial{\pmb \beta}$
and $\Sigma$ be the variance-covariance matrix of the sampling distribution of $n^{1/2} \tilde U_n (\pmb \beta)$.  
The following theorem presents the asymptotic  properties of the maximum pseudo-partial likelihood estimator.
\begin{theorem}
\label{thm 3.1}
Suppose that conditions C1-C3   in the 
Appendix   hold.  For each positive integer $L$, 
as $n \to \infty$, $\hat {\pmb \beta}_L$     
converges to the true parameter ${\pmb \beta}_0$ in probability.   Furthermore, 
$n^{1/2}(\hat{\pmb \beta}_L - {\pmb \beta}_0)$ converges weakly to a    mean  zero  Gaussian  distribution  with
covariance matrix  
$\Psi ({{\pmb \beta}_0}) 
= \Gamma^{ - 1}{({{\pmb \beta} _0})} 
\Sigma \Gamma^{ - 1}{({{\pmb \beta}_0})}$
where 
$\Gamma (\pmb \beta)= \mathop {\lim }\limits_{n \to \infty } \Gamma_n (\pmb \beta)$.  
\end{theorem}
The regularity conditions  and  the proof are  provided in  the Appendix.

\section{Numerical Studies}
\label{sec:verify}
\subsection{Simulation Study}
I   conducted  a simulation study  to evaluate   the finite  sample performance of the resulting pseudo-partial likehood 
(PPL)  estimates compared to   those of  the partial likelihood  (PL)  method  \citep{wangetal1993}.   
I  set the recruitment  time to  $\nu  = 10$ and generated  the onset times   from an exponential  distribution   with mean 1. 
The covariates ${\tilde Z}_1$  and ${\tilde Z}_2$ were generated from a standard normal distribution and  a  Bernoulli distribution 
with probability $p=P({\tilde Z}_1 > 0)$, respectively. Then the hazards of survival times were generated from  the Cox PH model as
$h_1(t|{\tilde Z}_1=z_1, {\tilde Z}_2=z_2)=2\exp(0.5z_1+z_2)$,  $h_2(t|{\tilde Z}_1=z_1, {\tilde Z}_2=z_2)=2t\exp(0.5z_1+z_2)$,
and $h_3(t|{\tilde Z}_1=z_1, {\tilde Z}_2=z_2)=3t^2\exp(0.5z_1+z_2)$  which  represent  constant, increasing and U-shape hazards, respectively. 
I considered the sample size of $n=200$ and $n=400$. The residual censoring times are generated from a uniform 
distribution on $(0,\theta)$ where $\theta$   yields   0\%, 20\%  or 40\%  rates of censoring. 
I repeated this simulation $N=1000$ times and summarized the results. \\
Table \ref{tab:1.3} reports the simulation results  from the estimation methods under the Cox PH  model with right-censored biased sampling data 
in terms of empirical  bias (Bias),  empirical standard deviation (ESD), and the averaged  robust standard error (ASE) of the estimates.
The results suggest that the PPL estimates are almost unbiased as those of the PL method. However, the PPL estimates are more efficient than those of
the PL method. This is because the PPL method takes  the information in the parametric distribution of the truncation distribution  into account. 
Also, the ESD and  the ASE  of the estimates  from the PPL method  are almost equal. These results remain the same in different scenarios including 
the type of hazards, the censoring rates, and sample size. It should be pointed out increasing the  censoring rates  resulted in an increase in the empirical 
bias and standard deviations of the estimates while increasing the sample size to $n=400$ has improved the results in terms of accuracy and precision.
%The simulation results improved in terms of accuracy and precision when the sample size increases to $n=400$. 
Nevertheless, the pattern of the results remained unchanged for all scenarios.  It is also noteworthy that the proposed method may become unstable 
for large censoring rates  due to unstable weight estimates. However, the method  outperforms the partial likelihood 
method for low and moderate rates of censoring.     
\begin{table}
\small
\caption{Simulation results on 
the estimation method  for  the Cox PH model with biased sampling data 
when    truncation times follow Exponential  distribution;  PL: partial likelihood; PPL: pseudo-partial likelihood; C: censoring rate; Bias: empirical Bias;
ESD: empirical standard deviation;  ASE: average standard error; $n_1=200$; $n_2=400$; $h_1$: constant baseline hazard;  
$h_2$: increasing baseline hazard; $h_3$: U-shape baseline hazard.}
\label{tab:1.3}
\centering
\begin{tabular}{|c|c|cc|ccc|}
\hline
              & C                  &         & PL &  & PPL &  \\ \cline{1-7}  
$h_1$    &  & Bias ($\beta_1$,$\beta_2$)  &    ESD ($\beta_1$,$\beta_2$)  &  Bias ($\beta_1$,$\beta_2$)  & ESD ($\beta_1$,$\beta_2$)  &  ASE ($\beta_1$,$\beta_2$)   \\  
\hline
              &   0\%         &   (0.001, 0.007) &   (0.084, 0.158)      & (-0.003, 0.001)  &  (0.067, 0.144) &  (0.062, 0.141)      \\
$n_1$    &  20\%        &  (-0.003, 0.008)  &  (0.093, 0.173)      & (-0.005, 0.005)  &  (0.078, 0.165) &  (0.072, 0.169)      \\
              &  40\%       & (0.005, 0.009)  &   (0.106, 0.194)        & (-0.008, -0.009)  &  (0.090, 0.187)  &  (0.088, 0.183)    \\
\hline
             &   0\%        &   (0.001, 0.005) &   (0.059, 0.138)       & (-0.001, 0.001)  &  (0.049, 0.122)  &   (0.045, 0.125)     \\
$n_2$   &  20\%       &  (0.002, 0.007)  &  (0.065, 0.157)       & (-0.003, 0.003)  & (0.055, 0.145)  &  (0.052, 0.153)     \\
             &  40\%       & (-0.004, 0.008)  &  (0.074, 0.173)         & (-0.006, -0.007)  &  (0.068, 0.160)  &  (0.076, 0.166)      \\
\hline
$h_2$   &   &    &&&    &\\ \cline{1-4} 
\hline
           &   0\%      &   (-0.004, 0.007)  &   (0.083, 0.163)    & (0.002, 0.002)  &  (0.074, 0.153) & (0.071, 0.157)      \\
$n_1$ &  20\%     &  (0.007, 0.005)  &   (0.090, 0.178)     & (0.005, 0.004)  &  (0.081, 0.169)  &  (0.085, 0.173)     \\
           &  40\%      & (-0.008, 0.004)  &  (0.104, 0.203)     & (-0.008, -0.007)  & (0.094, 0.182)   &  (0.103, 0.190)   \\
\hline 
           & 0\%        &   (-0.003, 0.017)  &   (0.075, 0.158)       & (0.001, -0.002)  &  (0.065, 0.139)  &  (0.075, 0.144)      \\
$n_2$ & 20\%      &  (-0.005, 0.020)  &  (0.084, 0.165)       & (0.004,  0.003)   &  (0.073, 0.147)  &  (0.081, 0.152)     \\
           & 40\%       &   (0.008, 0.021)  &  (0.091, 0.184)       & (-0.006, -0.005)  &  (0.088, 0.175)   &  (0.094, 0.181)      \\
\hline
$h_3$ &   &            &&    &     &\\ \cline{1-4} 
\hline
            &  0\%      &   (0.005, 0.003)   &   (0.082, 0.161)      & (0.002, -0.003)  &  (0.078, 0.143)  &   (0.075, 0.147)    \\
$n_1$  & 20\%      &  (-0.007, 0.004)    &   (0.091, 0.179)      & (0.004, 0.005)  & (0.086, 0.154)  &  (0.088, 0.161)     \\
            & 40\%     &   (0.008, 0.008)    &   (0.105, 0.206)       & (-0.005, -0.006)  &  (0.099, 0.193)   &   (0.105, 0.198)    \\
\hline
           & 0\%          &   (0.002, 0.001)  &   (0.057,  0.113)    &   (-0.001, 0.001)  & (0.057, 0.102)  &   (0.053, 0.104)    \\
$n_2$ & 20\%         &  (-0.004, 0.002)  &  (0.063, 0.125)     &  (0.004, -0.002)   &  (0.063, 0.116)  &  (0.060, 0.113)       \\
           & 40\%         & (-0.006, 0.005)    &   (0.073, 0.144)     & (-0.007, -0.005)  &  (0.073, 0.128)  &   (0.076, 0.133)     \\
\hline
\end{tabular}
\end{table}

\subsection{\large{HIV-infection and AIDS}}   
I employ the  pseudo-partial likelihood method  to estimate the regression coefficients in the Cox PH model  for a set of prevalent cohort data
from the Amsterdam Cohort Study on HIV infection and AIDS  \citep{geskus2000}. The data were collected among   homosexual men 
who had experienced HIV-infection prior to the study recruitment,  but none developed AIDS as discussed in \citet{geskus2000}.
While for those who are prospectively identified, the midpoint of the date of the last seronegative test and the  first seropositive 
test was considered  as the date of seroconversion. For seroprevalent cases, the situation is different. \citet{geskus2000} has thoroughly studied 
this issue and presented a marker-based approach, using CD4 counts, for imputing missing dates of seroconversion for such  cases.  \\ 
The data set consist of  $n=204$ prevalent cases  who have been infected by HIV before the recruitment time.  
The survival  time the time from HIV-infection to AIDS, the left-truncation  variable 
is the time between HIV-infection and study recruitment.   The covariates   are  age at HIV infection and a dichotomous variable 
CCR5 (C-C motif chemokine receptor 5) genotype with levels WW  (wild type allele on both chromosomes) and  WM (mutant allele on one chromosome). 
Among 204 patients, the survival time  of 57 subjects (28\%) were right-censored by the end of study.
The purpose  is to study the impact of the covariates  at HIV infection on the survival times under the assumption that their hazard follows a Cox PH model.  \\
I have estimated the truncation distribution using  the nonparametric method (the conditional MLE) of \citet{wang1991}. 
Figure \ref{fig: 3} depicts this estimator (red curve) and indicates an exponential trend between 1.5 and 2.5 years. 
Following  \citet{rabhi2020}, the Weibull distribution  with the shape parameter value  4.80  and scale parameter value 2.04  
(blue dashed) seems to support  the nonparametric estimator of the truncation distribution (red line).  \\
Table \ref{tab: 2.3} reports the estimates , their  standard errors of the effects of age and CCR5, and their p-values under the Cox PH model  for the HIV/AIDS data 
using the proposed pseudo-partial (PPL) method and the partial likelihood (PL) method \citep{wangetal1993}.  Both methods suggest that there is strong evidence 
that the patients' age at HIV infection  and WW  genotype  are positively  associated with the survival times at 5\% significance level.   
\begin{table}
\small
\caption{Estimated regression coefficients  and their  standard errors under the Cox PH model  for the HIV/AIDS data;   
PL: partial likelihood; PPL: pseudo-partial likelihood; Est: estimated regression coefficients;  SE: standard error.}
\label{tab: 2.3}
\label{fig: 3}
\centering
\begin{tabular}{|c|cc|c|}
\hline
                 & {AGE}           &    {CCR5}  &      \\ \cline{1-4}  
Methods   & {Est} (SE)     &{Est} (SE)    &    P-Value  (AGE, CCR5)      \\    %  95\% CI \hspace{0.5cm}  95\% CI 
\hline
PL            & 0.041 (0.013)      &  0.892  (0.233)  &  (0.002, 0.000)     \\   %   (???, ???) \hspace{0.25cm} (???, ???)
PPL          &   0.022 (0.010)     &  0.673 (0.228)   & (0.027, 0.003)  \\  %  (???,  ???)  \hspace{0.25cm}  (???, ???)
\hline
\end{tabular}
\end{table}

\begin{figure}[htb]
\centering
\includegraphics[scale=0.4]{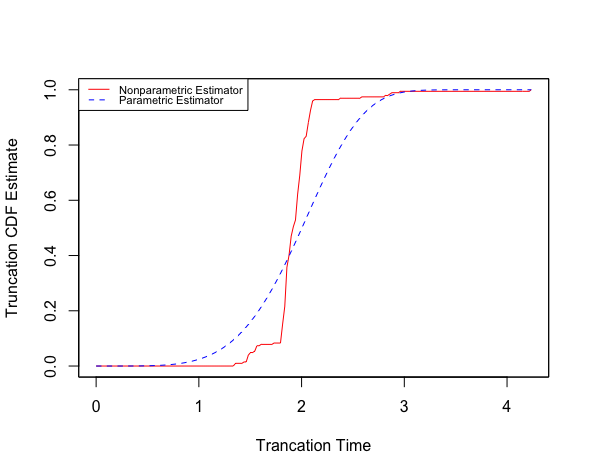}
\caption{Nonparametric estimator (solid line) and parametric estimator based on  Weibull family (dashed line).}
\end{figure}

\section{Discussion}
\label{sec:conc}
In this  article, I  proposed a pseudo-partial likelihood method for estimating parameters in the Cox PH model for right-censored biased 
sampling data via risk set sampling by correcting the bias of sample  risk sets. I further studied the asymptotic 
properties of the resulting estimator.  The simulation study showed that the resulting estimator outperform the maximum partial likelihood estimator 
when the truncation distribution is known.  I also used the proposed method to analyze   a set of   HIV/AIDS data. \\
Although  the   residual censoring  distribution was assumed to be  independent of covariates, the   methods 
can be adapted to account for   covariate-dependent censoring.   If the censoring distribution depends 
on covariates,  then  the function $\Omega_C(t)$ needs to be replaced    by  $\Omega_C(t|\textbf z) = \int_0^t S_C(u| \textbf z) du$, respectively. 
When covariates are discrete,  the conditional weight functions  can be consistently estimated by   
${\hat \Omega}_C(t| \textbf z) = \int_0^t {\hat S}_C( u| \textbf z) g_{\theta} (u) du$,  where ${\hat S}_C(t| \textbf z)$   
is obtained  by  the Kaplan-Meier estimator.  For continuous covariates, we can either 
consider a regression model for   the censoring distribution or adopt the local Kaplan-Meier estimator  \citep{wangwang2014}.  
A careful   investigation on these topics   is warranted. 

\section{Appendix}
\label{app: A0}
We assume the following regularity conditions to establish   the asymptotic properties of the estimator.
\begin{enumerate} [C1.]
\item   The covariate vector ${\textbf Z}$ is a bounded random vector  and  ${\pmb \beta}_0$ lies in a compact set $\mathcal B$.
\item  $P(C>0) > 0$.
\item $S_U(t)$ and $S_C(t)$  are absolutely continuous for $t \in [0, \tau]$ where $\tau = \sup\left\{t: P(Y \ge t ) > 0 \right\}$.
\item  $\Gamma ({\pmb \beta}_0)$ is   positive definite    where $\Gamma (\pmb \beta) = \mathop {\lim }\limits_{n \to \infty } {\Gamma _{n}}(\pmb \beta)$ 
and $\Gamma_{n}(\pmb \beta)= - \partial U_{n}(\pmb \beta )/\partial {\pmb \beta}$.
\end{enumerate}
%\numberwithin{equation}{subsection}
%\numberwithin{corollary}{subsection}
%\numberwithin{lemma}{subsection}

%\subsection{Asymptotic Properties}
\section*{Proof of Theorem \ref{thm 3.1} }
\label{app: A2}
To show the consistency,  we consider  the following log pseudo-partial likelihood function. 
$$
K_n(\pmb \beta) = \frac{1}{n}\sum\limits_{i = 1}^n {\int_0^\tau  {\left[ {{{({\pmb \beta} - {{\pmb \beta}_0})'}}{{\textbf z}_i}
- \log \left\{ {\frac{{S_{lB} ^{(0)}({\pmb \beta} ,t)}}{{S_{lB}^{(0)}({\pmb \beta}_0 ,t)}}} \right\}} \right]} } dN_i(t).$$
Note that $\Gamma_{n}(\pmb \beta)= \frac{\partial^2 K_n(\pmb \beta)}{\partial {\pmb \beta} \partial {\pmb \beta}}$.
Further, for any ${\pmb \beta} \in \mathcal B$, as $n \to \infty$,  $\Gamma_{n}(\pmb \beta)$ converges almost surely  to $\Gamma$   
which is assumed to be positive definite.  This implies that  $K_{n}({\pmb \beta})$  is a concave function of $\pmb \beta$.   Therefore,  
the unique maximizer of $K_{n}({\pmb \beta})$ converges in probability to the unique maximizer of $K({\pmb \beta})$ 
\citep{andersengill1982} or  $\hat{\pmb \beta}$ converges in probability to  ${\pmb \beta}_0$.  
This completes the proof of consistency.  \\  
For  the  proof of  asymptotic normality,  we first define    
$${S^{(k)}}({\pmb \beta },y) = n^{-1}\sum\limits_{j=1}^{n} {\textbf z}_j^{ \otimes k}\exp ({{\pmb \beta} ^T}{\textbf z_j})
{\{ {\Omega _C}(y)/{\Omega _C}({y_j})\}{\zeta_j}(y)}, \quad  k=0,1,2.$$  
Also,   $n^{1/2}{\tilde V}_{n}(\pmb\beta)$  defined in \eqref{eq: 15}  can be expanded as 
$$
{n^{ - 1/2}}\sum\limits_{i = 1}^n {\int_0^\tau {\left\{ {{\textbf z_i} - \frac{{S^{(1)}(\pmb\beta ,t)}}{{S^{(0)}(\pmb\beta ,t)}}} \right\}dN_i(t)} }  
+ {n^{ - 1/2}}\sum\limits_{i = 1}^n {\int_0^\tau  {\left\{ {\frac{{\tilde S^{(1)}(\pmb\beta ,t)}}{{\tilde S^{(0)}(\pmb\beta ,t)}} 
- \frac{{S^{(1)}(\pmb\beta ,t)}}{{S^{(0)}(\pmb\beta ,t)}}} \right\}dN_i(t)} }.  
$$
From Remark  \ref{rem: 1}, $n^{1/2}{\tilde U}_{n}(\pmb\beta)$ and $n^{1/2}{\tilde V}_{n}(\pmb\beta)$ are asymptotically equivalent 
(see the proof in the Appendix). Moreover,  
\begin{align}
\label{eq: B2}
n^{1/2}{\tilde V}_{n}(\pmb\beta)= \frac{1}{{\sqrt n }}\sum\limits_{i = 1}^n {\int_0^\tau  {\left\{ {{\textbf z_i} 
- \frac{{s^{(1)}(\pmb\beta ,t)}}{{s^{(1)}(\pmb\beta ,t)}}} \right\}dM_{i}(t)} }  
+ \frac{1}{{\sqrt n }}\sum\limits_{i = 1}^n {\int_0^\tau  {G(\pmb\beta ,t)\frac{{d{M_{Ci}}(t)}}{{\pi (t)}}} }  + o_P(1),
\end{align}
where, if  we let $H_C(t)$ be the hazard function for the residual censoring, then 
$${{M}_{Ci}}(t) = I({V_i} \le t,{\delta _i} = 0) - \int_0^t {I({V_i} \ge u)d{{H}_C}} (u).$$
In addition, if we define $\pi(t)= S_C(t)S_V(t)$ and   further ${{h}_k}(t) = I(t \le {Y_k})\int_t^{{Y_k}} {{{S}_C}} (u)du$, then we have 
$$G(\pmb\beta ,t) = \mathop {\lim }\limits_{n \to \infty } \frac{1}{{{n^2}}}\sum\limits_{i = 1}^n 
{\sum\limits_{k = 1}^n {\frac{{ {\zeta_k}({y_i}){\Omega}_C(Y_i){{\textbf Z}_k}\exp ({{\pmb\beta'}}{{\textbf z}_k})\left[ 
{1/{{\Omega }_C}({y_k})} \right]^2{{h}_k}(t)}}{{S^{(0)}(\pmb\beta ,{Y_i})}}} }.$$
Note that  $(\hat\Omega_C(Y_k)- \Omega_C(Y_k))$ can be expressed as   sum of  independent and identically distributed  (i.i.d) 
martingale processes \citep{pepe1991} as below. 
$$
{n^{1/2}}({\hat \Omega _C}({Y_K}) - {\Omega _C}({Y_K})) = {n^{ - 1/2}}\sum\limits_{j = 1}^n 
{\int_0^\tau  {{h_k}(t)\frac{{d{M_{Cj}}(t)}}{{\pi (t)}}} }.
$$
By a similar  argument to that of \citet{qinshen2010},  the  terms in equation \eqref{eq: B2} 
are summation of   i.i.d zero-mean martingale processes.  
Hence,  under  the regularity conditions C1-C3, by the martingale central limit theorem,  
$n^{1/2}{\tilde V}_{n}(\pmb\beta)$ converges  weakly to a zero-mean Gaussian process with covariance matrix 
$$
\Sigma = 
E\left[ {\int_0^\tau  {{\left( {{\textbf Z_i} - \frac{{s^{(1)}(\pmb\beta ,t)}}{{s^{(0)}(\pmb\beta ,t)}}} \right)dM_{i}(t)   
+ \frac{{G(\pmb\beta ,t)}}{{\pi (t)}}d{M_{Ci}}(t)}}} \right]^{ \otimes 2}.
$$
Let $\hat{\pmb \beta}$ be  a solution to  ${\tilde V}_{n}(\pmb\beta)=\textbf 0$. By Taylor expansion, we have   
$$n^{1/2} (\hat{\pmb \beta}- {\pmb \beta}_0)= \{ \Gamma_{n} ({\pmb \beta}^*) \}^{-1} n^{1/2}{\tilde V}_{n}(\pmb\beta) +o_P(1),$$
where ${\pmb \beta}^*$ is on the line segment between $\hat{\pmb \beta}$ and ${\pmb \beta}_0$. 
Since $\hat{\pmb \beta}$ is consistent for ${\pmb \beta}_0$ and $\Gamma$ is continuous at ${\pmb \beta}_0$,  
we see that $\Gamma_{n}(\pmb \beta)$ and as such  $\Gamma_{n}({\pmb \beta}^*)$ converge to $\Gamma$ almost surely. 
Therefore, by  Slutsky's theorem, $\sqrt n (\hat{\pmb \beta} -{\pmb \beta}_0)$ converges to a zero-mean Gaussian distribution 
with  covariance matrix   $\Psi ({{\pmb \beta}_0})= \Gamma^{ - 1} ({{\pmb \beta}_0}) \Sigma \Gamma^{ - 1} ({{\pmb \beta}_0})$.   
This completes the proof.
\begin{remark}
The Hessian matrix   $\Gamma (\pmb \beta)$   and the covariance matrix $\Sigma$   can
be consistently  estimated by $\widehat\Gamma_{n}$  and $\widehat \Sigma$, respectively,  
where $\widehat\Gamma_{n}=\Gamma_{n}(\hat{\pmb \beta})$   and
$$ 
\widehat { \Sigma} = {n^{ - 1}}
\sum\limits_{i = 1}^n   {\int_0^\tau  {\left\{ {\left( {{\textbf z_i}  %  {\left[
- \frac{{\tilde S^{(1)}(\hat {\pmb\beta} ,t)}}{{\tilde S^{(0)}(\hat {\pmb\beta} ,t)}}} \right)d\widehat M_{i}(t) 
+ \frac{{\widehat G(t)}}{{\bar y(t)}}d{{\widehat M}_{Ci}}(t)}  \right \}^{ \otimes 2} }},
$$
where  ${{\widehat M}_{Ci}}(t) = I({y_i} \le t,{\delta _i} = 0) - \int_0^t {I({y_i} \ge u)d{{\widehat H}_C}} (u)$, 
$\bar y(t) = {n^{ - 1}}\sum\limits_{i = 1}^n {I(\min ({c_i},{v_i}) > t)}$,  \\
and  ${{\hat h}_k}(t) = I(t \le {y_k})\int_t^{{y_k}} {{{\hat S}_C}} (u)du$. In addition, $\widehat H_C(t)$ is the Nelson-Aalan estimator for  the residual censoring time, and 
$$\widehat G(t) = \mathop {\lim }\limits_{n \to \infty } \frac{1}{{{n^2}}}\sum\limits_{i = 1}^n 
{\sum\limits_{k = 1}^n {\frac{{ {\zeta_i}({y_i}){\hat\Omega}_C(y_i){{\textbf z}_k}\exp ({{\hat {\pmb\beta'} }}{{\textbf z}_k})\left[ 
{1/{{\hat \Omega }_C}({y_k})} \right]^2{{\hat h}_k}(t)}}{{\tilde S^{(0)}(\hat {\pmb\beta} ,{y_i})}}} }$$
The covariance matrix  $\Psi({{\pmb \beta}})$ can then be consistently  estimated by
$\widehat\Gamma_{n}^{ - 1}  {\widehat \Sigma} \widehat\Gamma_{n}^{ - 1}$. 
\end{remark}

\section*{Proof of Remark \ref{rem: 1} }     
Let $F^u(t)=P(Y \le t, \Delta=1)$  be  the distribution functions of  uncensored  data. The proof is carried out similar to that of  \citet{wang1996}. 
Since  $E\left[{B_{lj}}(t)|Y_j \right]= \left[\Omega(t)/\Omega_C(Y_j)\right]{\zeta _j}(t)$, we have   
$$E\left[S^{(k)}({\pmb \beta},t)\right]=E\left\{ E\left[S_{lB}^{(k)}({\pmb \beta},t) | {\textbf Z}_j,Y_j \right]\right\}= s^{(k)}({\pmb \beta},t)$$
The regression coefficients in the Cox PH model can be estimated  by  replacing  ${\xi_j}(Y_i)$ in  $U_n ({\pmb \beta})$
with their expected value given $(Y_i,Y_j)$ which  yields  the below  weighted estimating function.  
\begin{align}
{V_n (\pmb \beta)}=\frac{1}{n}\sum\limits_{i = 1}^n {V_{i}}(\pmb \beta)=
\frac{1}{n}\sum\limits_{i = 1}^n {\delta_i
\left[ {{\textbf z_i} - \frac{{S^{(1)}}({\pmb \beta },y_i)}{{S^{(0)}}({\pmb \beta },y_i)}} \right]}.
\end{align}
To show the asymptotic equivalency, I expand  ${n^{1/2}}U_n (\pmb \beta)$   as
\begin{align}
\label{eq :1.0.3}
 {n^{ - 1/2}}\sum\limits_{i = 1}^n {{\delta _i}\left[ {{{\textbf z}_i} - \frac{{{s^{(1)}}(\pmb \beta ,{y_i})}}{{{s^{(0)}}(\pmb \beta ,{y_i})}}} \right]}
- {n^{ - 1/2}}\int_0^\tau {\left[ {\frac{{S_{lB}^{(1)}(\pmb \beta ,{t})}}{{S_{lB}^{(0)}(\pmb \beta ,{t})}}
- \frac{{{s^{(1)}}(\pmb \beta ,t)}}{{{s^{(0)}}(\pmb \beta ,t)}}} \right]} d\hat F^u(t),   
\end{align}    
where $\hat F^u$ is the empirical distribution function of $F^u$.    Applying Taylor expansion,  we  obtain  
$$
{n^{1/2}}S_{lB}^{(1)}(\pmb \beta ,t)\left[ {S_{lB}^{(0)}(\pmb \beta ,t) - {s^{(0)}}(\pmb \beta ,t)} \right]
= {n^{1/2}}{s^{(1)}}(\pmb \beta ,t)\left[ {S_{lB}^{(0)}(\pmb \beta, t) - {s^{(0)}}(\pmb \beta ,t)} \right] + {o_P}(1).
$$
The second term in \eqref{eq :1.0.3} can be further expressed  as
$$
- {n^{ - 1/2}}\sum\limits_{i = 1}^n {\int_0^\tau {\frac{{{B_{li}}(t)\exp ({{{\pmb \beta'}}{{\textbf z}_i}})}}{{{s^{(0)}}(\pmb \beta ,t)}}} }
\left[ {{{\textbf z}_i} - \frac{{{s^{(1)}}(\pmb \beta ,t)}}{{{s^{(0)}}(\pmb \beta ,t)}}} \right]d\hat F^u(t) + {o_P}(1).
$$  
Similarly,  ${n^{1/2}}{V_n(\pmb \beta)}$  can be expanded as   
\begin{align}
\label{eq: 1.0.4}
{n^{ - 1/2}}\sum\limits_{i = 1}^n {{\delta_i}\left[ {{{\textbf z}_i}- \frac{{{s^{(1)}}(\pmb \beta ,{y_i})}}{{{s^{(0)}}(\pmb \beta ,{y_i})}}} \right]}
- {n^{ - 1/2}}\int_0^\tau {\left[ {\frac{{S^{(1)}(\pmb \beta ,{t})}}{{S^{(0)}(\pmb \beta ,{t})}}
- \frac{{{s^{(1)}}(\pmb \beta ,t)}}{{{s^{(0)}}(\pmb \beta ,t)}}} \right]}d\hat F^u(t).
\end{align}
The second term in  \eqref{eq: 1.0.4} is asymptotically equivalent to
$$
- {n^{ - 1/2}}\sum\limits_{i = 1}^n {\int_0^\tau {\frac{{\left\{ {{\Omega _C}(t)/{\Omega _C}({y_i})}
\right\}{\zeta_i}(t)\exp ({{\pmb \beta'}{{\textbf z}_i}})}}{{{s^{(0)}}(\pmb \beta ,t)}}} } \left[ {{{\textbf z}_i}
- \frac{{{s^{(1)}}(\pmb \beta ,t)}}{{{s^{(0)}}(\pmb \beta ,t)}}} \right]d\hat F^u(t) + {o_P}(1).
$$                  
Thus,  the difference of the two estimating functions can be expressed as 
\begin{align}
 & {n^{ - 1}}\sum\limits_{i = 1}^n {{U_{li}}}(\pmb \beta)  - {n^{ - 1}}\sum\limits_{i = 1}^n {V_{i}}(\pmb \beta)  \nonumber \\
 & ={n^{ - 1}}\int_0^\tau {\left\{ {\frac{{{S^{(1)}}({\pmb \beta} ,t)}}{{{S^{(0)}}({\pmb \beta} ,t)}}
- \frac{{{s^{(1)}}({\pmb \beta} ,t)}}{{{s^{(0)}}({\pmb \beta} ,t)}}} \right\}d\hat F^u(t)}                                                
- {n^{ - 1}}\int_0^\tau {\left\{ {\frac{{S_{lB}^{(1)}({\pmb \beta} ,t)}}{{S_{lB}^{(0)}({\pmb \beta} ,t)}}
- \frac{{{s^{(1)}}({\pmb \beta} ,t)}}{{{s^{(0)}}({\pmb \beta} ,t)}}} \right\}d\hat F^u(t)}  \nonumber \\                          
 &= {n^{ - 1}}\sum\limits_{i = 1}^n {\int_0^{\tau} {\frac{{\left[ {\left\{ {{\Omega _C}(t)/{\Omega _C}({y_i})} \right\}{\zeta_i}(t)
- {B_{li}}(t)} \right]\exp ({\pmb \beta' }{{\textbf z}_i})}}{{{s^{(0)}}({\pmb \beta} ,t)}}} } \left\{ {{{\textbf z}_i}
- \frac{{{s^{(1)}}({\pmb \beta} ,t)}}{{{s^{(0)}}({\pmb \beta} ,t)}}} \right\}d\hat F^u(t)+ {o_P}(1) \nonumber \\               
 &= {n^{ - 2}}\sum\limits_{i = 1}^n {\sum\limits_{j = 1}^n \frac{{\left[ {\left\{ {{\Omega _C}({y_j})/{\Omega _C}({y_i})} \right\}{\zeta_i}({y_j})
- {B_{li}}({y_j})} \right]\exp({{\pmb \beta'}{{\textbf z}_i}})}}{{{s^{(0)}}(\pmb \beta ,{y_j})}}\left[ {{{\textbf z}_i}
- \frac{{{s^{(1)}}(\pmb \beta ,{y_j})}}{{{s^{(0)}}(\pmb \beta ,{y_j})}}} \right] } + {o_P}(1)  \nonumber \\
&= {n^{ - 2}}\sum\limits_{i = 1}^n {\sum\limits_{j = 1}^n {{\mathcal D_{ij}}} }  + {o_P}(1),
\end{align}
where
\begin{align}
{\mathcal D_{lij}} &= \frac{1}{2}\frac{{\left[ {\left\{ {{\Omega _C}({y_j})/{\Omega _C}({y_i})} \right\}{\zeta_i}({y_j})
- {B_{li}}({y_j})} \right]\exp({{\pmb \beta'}{{\textbf z}_i}})}}{{{s^{(0)}}(\pmb \beta ,{y_j})}}\left[ {{{\textbf z}_i}
- \frac{{{s^{(1)}}(\pmb \beta ,{y_j})}}{{{s^{(0)}}(\pmb \beta ,{y_j})}}} \right] \nonumber  \\
 & + \frac{1}{2}\frac{{\left[ {\left\{ {{\Omega _C}({y_i})/{\Omega _C}({y_j})} \right\}{\zeta_j}({y_i})
- {B_{lj}}({y_i})} \right]\exp ({{{\pmb\beta'}}{{\textbf z}_j}} )}}{{{s^{(0)}}(\pmb \beta ,{y_i})}}\left[ {{{\textbf z}_j}
- \frac{{{s^{(1)}}(\pmb \beta ,{y_i})}}{{{s^{(0)}}(\pmb \beta ,{y_i})}}} \right]. \nonumber
\end{align}
Using the variance calculation techniques of $U$ and $V$-statistics  \citep[page 183, Lemma A (iii)]{serfling2009}, we have
$$\text{Var}({n^{ - 1}}\sum\limits_{i = 1}^n {{U_{lid}}} - {n^{ - 1}}\sum\limits_{i = 1}^n {{V_{id}}} )
= \frac{{4{\tau _{1l}}}}{n} + O({n^{ - 2}})  \quad \textrm{for \ d=1,...,p},$$
where ${\tau _{1l}} = \text{cov}({\mathcal D_{12d}},{\mathcal D_{13d}})$. 
It follows from the definition of $B_{lj}(Y_j)$ that  conditional on $({\textbf Z}_1,Y_1),({\textbf Z}_2,Y_2),({\textbf Z}_3,Y_3)$,  
the random vectors $\mathcal D_{12}$ and  $\mathcal D_{13}$  are  independent  and  have   zero expectations. Hence, 
$$
{\tau _{1d}} = E\left\{ {E\left[ {{\mathcal D_{l12d}}\left| {({{\textbf Z}_1},{Y_1}),({{\textbf Z}_2},{Y_2}),({{\textbf Z}_3},{Y_3})} \right.}
\right]\left[ {{\mathcal D_{l13d}}\left| {\left| {({{\textbf Z}_1},{Y_1}),({{\textbf Z}_2},{Y_2}),({{\textbf Z}_3},{Y_3})} \right.} \right.} \right]} \right\} = 0,
$$
and 
$$
\text{Var}({n^{- 1}}\sum\limits_{i = 1}^n {{U_{lid}}} - {n^{ - 1}}\sum\limits_{i = 1}^n {{V_{id}}})=O({n^{ - 2}}).
$$
This implies  that  ${n^{1/2}}U_n (\pmb \beta)$ and  ${n^{1/2}}{V_n (\pmb \beta)}$      are asymptotically equivalent.

%\newpage

%===============================================================================================
%===============================================================================================
%===============================================================================================
%===============================================================================================
%============================== References =======================================================
%===============================================================================================
%===============================================================================================
%===============================================================================================

\bibliographystyle{apalike}

\bibliography{reference}

\end{document}